# From Smart to Intelligent Utility Meters in Natural Gas Distribution Networks

Ghaith Matalkah
School of Electrical and
Computer Engineering
Georgia Institute of Technology
Atlanta, GA, USA
ghaith@gatech.edu

Edward J. Coyle
School of Electrical and
Computer Engineering
Georgia Institute of Technology
Atlanta, GA, USA
ejc@gatech.edu

*Abstract*—We propose a novel method for monitoring gas distribution networks (GDNs) using intelligent sensor nodes that can be integrated with existing smart gas meters (intelligent meters). The method aims at detecting and locating gas leaks in GDNs in real time. The intelligent meters leverage wireless connectivity in existing smart meters to collaborate in implementing this method, which comprises an active acoustic probing phase and a passive linear imaging phase. In the active acoustic phase, the intelligent meters collaboratively discover the topology of the monitored pipeline network using a novel acoustic pulse reflectometry technique. In the passive linear imaging phase, the intelligent meters use their knowledge of the pipeline network topology to create a linear image of the pipeline using the Time-Exposure Acoustic (TEA) algorithm. The resulting image reveals the presence and locations of active gas leaks in the network. We present the theoretical basis of the method and show results of implementing it on experimental data collected in the lab.

## I. INTRODUCTION

The adoption of smart utility meters in the electric, natural gas, and water utility industry is a prominent example of an IoT application that has revolutionized an entire industry. Smart utility meters are deployed at utility consumption points to collect and aggregate real-time consumption data and communicate it back to utility servers for analysis and insight. The data communication part is done through data networks formed by the smart meters and other communication devices as part of a whole system known as Advanced Metering Infrastructure (AMI). In the natural gas utility industry, data collected by smart gas meters have enabled a host of new applications including automated meter reading, remote detection of tampering and gas theft, meter health reporting, and real-time consumption reporting to utility companies and customers [1]. These applications improved the gas utility infrastructure and eliminated the costs and inefficiencies associated with accomplishing the same tasks manually. However, despite the benefits of real-time data analytics brought by smart meters and AMIs, the natural gas utility infrastructure has not reaped the full benefits of IoT technologies in solving some of its lingering problems. One of those problems is leakage in gas transmission and distribution pipelines. Leak-related incidents resulting in the loss of human lives and economic damage continue to be reported worldwide. For example, in the United States, natural gas leak incidents cause on average 17 fatalities, 68 injuries, and $133 M in property damage every year [2]. Furthermore, natural gas is mainly composed of methane ($CH_4$), which is a greenhouse gas with global warming potential 28 to 36 times that of carbon dioxide [3] and its unnecessary release into the atmosphere through unrepaired leaks can exacerbate the ongoing problem of global warming.

Leaks in natural gas pipelines occur as a result of structural defects in pipelines. Such defects are mainly attributed to pipe corrosion, installation faults, and digging activities near pipelines. While many pipeline monitoring and leak detection solutions have been developed in the literature [4], most of that effort was focused on monitoring gas *transmission pipelines*, which are the upper tier of pipelines in the natural gas delivery system that deliver gas from upstream production sites to downstream storage site. *Distribution pipelines*, on the other hand, which are the lowest tier of pipelines that distribute natural gas from storage sites to residential and commercial destinations, have rarely been the target of those research efforts. The lack of efficient pipeline monitoring techniques for distribution pipelines allows leaks to go undetected for long periods of time, increasing their potential hazard. As existing gas distribution pipeline networks age, they become more susceptible to leaks and defects as it is evident in the decades-old gas distribution networks serving older cities in the United States and Europe. For instance, Figure 1 shows the levels of leaked methane gas, the main component of natural gas, as measured on the streets of Boston, MA in a study conducted in [5].

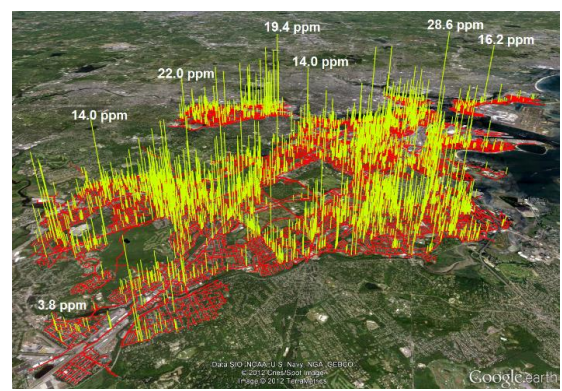

**Figure 1 - Levels of leaked Methane on the streets of Boston, MA. (with permission form Nathan Philips [5])**

Another study in [6] showed comparable levels of leaked natural gas on the streets of Washington, DC.

In this paper, we proposed a novel IoT-based method for monitoring Gas Distribution pipeline Networks (GDNs). The method is implemented using *intelligent meters*, which are intelligent wireless sensor nodes that are integrated with existing smart gas meters. Specifically, we propose a real-time fully automated monitoring method where intelligent meters leverage wireless connectivity in existing utility AMI to collaborate in implementing distributed acoustic-based signal processing and machine learning techniques aimed at detecting and locating leaks in GDNs. The intelligent meters collaboratively "learn" the structure of the pipeline network using a novel distributed active acoustic probing technique and generate linear acoustic images of the pipeline that reveal the presence and locations of leaks. Moreover, the leak detection techniques proposed here are designed with careful consideration to the practical constraints of GDNs, which prevent existing leak detection methods from being applied in GDNs. To the best of our knowledge, no similar leak detection method have been proposed or suggested in the literature. In the next section, we discuss existing acoustic-based gas leak detection methods and we describe our method in details along with experimental results in Section III.

## II. ACOUSTIC-BASED LEAK DETECITON IN NATURAL GAS PIPELINE NETWORKS

A wide variety of gas leak detection methods have been proposed in the literature using different sensing technologies such as optical and thermal imaging, chemical and bio-sensing, flow measurements, and acoustic sensing [4]. Acoustic-based methods were shown to have the capability of detecting and locating leaks from long distances with superior accuracy [7][8]. They also outperform methods based on other sensing technologies in speed of leak detection and the ability to be retrofitted to existing pipeline infrastructure [4][9]. This makes them an appropriate choice for application in GDNs since such networks are mostly buried underground and physical access to their pipelines is typically limited to the locations of utility meters.

In general, there are two categories of acoustic-based gas leak detection methods. The first category includes passive acoustic methods, which are based on detecting acoustic emission waves generated by active leaks in pipelines using sensors that continuously measure acoustic activity in or around pipelines. While methods in this category are similar in principle, different implementation in terms of signal processing and classification exist. For example, in [7] gas leaks signals were modeled as white noise signals passing through a spectral shaping filter. The filter coefficients are estimated using acoustic measurements from equidistant sensors that are deployed along the monitored pipeline. The filter coefficients are then used in deriving LPC Cepstrum coefficients, which are used as input features in to a KNN-based leak classifier. When a leak is detected, its location is determined based on comparing sensor signal amplitudes to a predetermined signal attenuation model. In [10], leaks were detected based on a correlation measure between sensor signals to leak signatures that are predetermined during a training phase and leaks were located using a method similar to that in [7].

The use of sensor signal cross-correlation to locate leaks has been used in several passive acoustic methods [11]. In [9] and [12], the Discrete Wavelet Transform (DWT) was used in denoising sensor signals prior to performing cross-correlation.

In [13], a hierarchical Wireless Sensor Network (WSN)-based method was proposed for monitoring pipeline networks using a passive acoustic method, where each sensor node uses DWT coefficients of recorded acoustic signals as input features into a support vector machine (SVM)-based leak classifier. Local node decisions are then aggregated in a sink node for final leak detection decision. Another WSN-based passive acoustic method was proposed in [14] and applied for detecting leaks in water and sewer networks. A major drawback of these passive methods is that they all require acoustic sensors to be coupled directly to the monitored pipeline along its extension, which impedes their application in underground gas distribution networks.

The second category is active acoustic methods, which are based on measuring the response of gas pipelines to acoustic excitation and detecting leak signature in the pipeline's acoustic response. In this category, an acoustic actuator (e.g., speaker, solenoid valve) is used to inject acoustic excitation waves into a gas pipeline at one end and the response of the pipeline is measured using a sensor (e.g., microphone, accelerometer) either at the same pipeline end (input response) or the other end (output response). The presence of leaks are detected as either reflections of the injected signal or as change in the acoustic impedance of the pipeline [15]. In [16], an active acoustic method was proposed in which a broadband pseudorandom acoustic signal was injected into one end of the pipeline and measured at both ends. The impulse response of the pipe was then estimated where leaks manifested as sharp pulses at time shifts corresponding to their locations. A different, and more practiced, approach in implementing active acoustic methods is using Acoustic Pulse Reflectometry (APR). In APR, an acoustic pulse in injected into a pipeline and its reflections are analyzed for detecting perforations in the pipeline walls. APR is based on the principle that when an acoustic wave propagating through a pipe encounters a sudden change in the cross-sectional area of the pipe, it experiences partial reflection and partial transmission at the boundary of the cross-sectional area change [17]. The cross-sectional area of a pipe $S$ is directly related the acoustic impedance of the medium inside it: $z = \rho c/S$, where $\rho$ is the density of the medium and $c$ is the speed of sound in the medium. Therefore, change in cross-sectional area constitutes change in acoustic impedance of the medium, which is a cause for wave reflection [17]. An incident wave with amplitude $P_0^+$ propagating inside a pipe through a change in the cross-section area from $S_0$ to $S_1$, generates a reflected wave with amplitude $P_0^-$ that is proportional to the size of the change [17]:

$$\frac{P_0^-}{P_0^+} = \frac{S_0 - S_1}{S_0 + S_1} \qquad (1)$$

Leak-causing defects in pipeline walls introduce abrupt changes in the cross-sectional area of a pipe, which cause acoustic reflections that can be measured in order to detect leaks. APR was reportedly first applied to leak detection in [18], where leaks were detected by comparing pulse reflection signals from a pipe under test to a reference signal collected from a similar leak-free pipe. In [15], the method was refined using an algorithm that reconstructs the bore profile of the pipe under test using the Input Impulse Response (IIR) estimated from APR measurements. In [19], a series of detailed experiments showed the ability of detecting leaks using APR in simple pipeline networks including L- and T-shape pipe junctions. In [8], experimental results showed that APR can be used to detect the presence of holes in wide-diameter pipes at distances up to 22 meters. One of the shortcomings of APR though is that small size leaks generate low SNR reflections that are hard to detect. This is can be inferred from equation (1). As a result, several methods aiming to enhance the SNR of leak reflections were proposed. In [20], the use of Maximum Length Sequence (MLS) as source signal and cross-correlation to estimate the IIR showed improved leak SNR. In [21], a chirp source signal was used and frequency-domain analysis of pulses in the IIR were applied to distinguish leaks from other reflection-causing features such as pipe junctions. Finally, in [22], orthogonal wavelet transform (OWT) was applied in de-noising pulse reflection signals and the complex Cepstrum analysis was used to detect reflections corresponding to leaks. Leak related peaks in the complex Cepstrum showed higher leak SNR.

APR-based methods are capable of detecting perforations and leaks in straight pipes. However, they perform very poorly in complex pipeline networks with features such as L-bends, cross- and T-junctions. In all of the works described above, only [19] have studied APR in simple pipeline networks and showed that only large perforation under certain conditions can be reliably detected. This is because pipeline joints and junctions also generate pulse reflections in the same mechanism that leaks do but are typically order of magnitude stronger [19]. Furthermore, acoustic waves propagating in complex pipelines exhibit high order reflections that destructively interfere with APR signals and make the extraction of leak signals nearly impossible. For these reason, such methods are not suitable for complex pipeline networks like GDNs. In the following section, we present a novel acoustic-based method that combines an improved version of APR with a passive acoustic technique for detecting and locating leaks reliably in GDNs.

## III. SYSTEM ARCHITECTURE AND OPERATION

As mentioned earlier, the monitoring method proposed here is intended for real-time detection and location of leaks in GDNs using intelligent meter nodes. Therefore, the design requirements of the method in terms of hardware and signal and data processing must accommodate the practical constraints of GDNs and the characteristics of leak signals occurring therein.

### A. Practical constraints of gas distribution networks

GDNs are intended to distribute gas from a single source (i.e., storage stations) to multiple destinations (i.e., homes and businesses) hence they tend to have tree-like topologies that exhibit high density of pipe branching. Moreover, GDN pipelines are completely buried underground except for utility meter locations at service points (i.e., GDNs terminals), which limits physical access to the pipelines *see Figure 2*. Consequently, the monitoring method must be able to detect and locate leaks in complex branching pipeline networks and do that strictly from the network terminals (i.e., meter locations). Another attribute of GDNs is that their pipeline topologies and gas flow patterns change over time as service points get on and off line periodically. Therefore, intelligent meters must be capable of dynamically discovering the topology of the monitored network.

### B. Properties of acoustic leak signals

The gas pressure in GDNs is typically less than 10 psi [23]. Therefore, active leaks have weak acoustic signal compared to leaks in transmission pipelines where gas pressure reaches up to 1500 psi. Moreover, the steady level of gas pressure in the pipeline creates a continuous leak signal with a waveform that depends on many factors including the shape and size of the leak, ambient temperature, and gas pressure [24]. Therefore, the only assumption we make about leak signal properties is they are continuous, random, and have low SNR.

### C. System Architecture

The monitoring method is implemented using a network of intelligent meter (IM) nodes, where IM nodes are divided into clusters. Each cluster includes a cluster head/gateway node and several IM nodes. IM nodes can communicate wirelessly either directly or indirectly through the cluster head node. Figure 2 shows a segment of a GDN monitored by a cluster of four IM nodes.

In addition to the utility meter module, each IM node is comprised of the following components: processing unit, wireless communication module, an acoustic actuator (e.g., speaker, pulse generator, or valve), and an acoustic sensor (e.g., microphone, accelerometer). We have used speakers and microphones in the experimental work presented here, therefore they are assumed to be the choice of acoustic transducers in the discussion below.

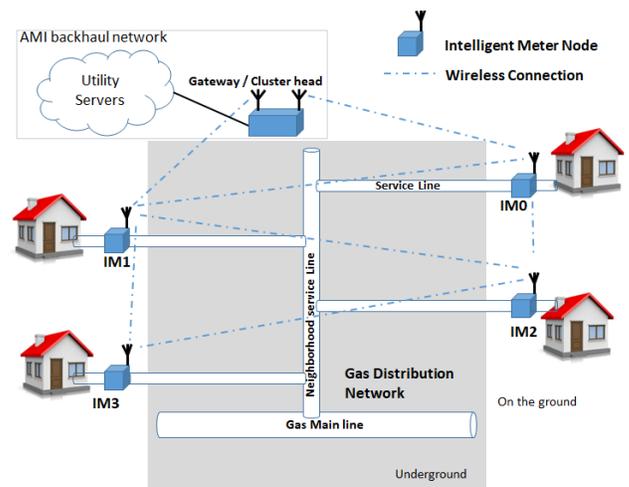

**Figure 2 – A segment of GDN monitored by intelligent meters. Gray-shaded area represents part of the pipeline network buried underground.**

### D. Leak Detction Algorithm

The leak detection and locating algorithm implemented by the IM nodes is comprised of two phases: an active acoustic probing phase, and a passive acoustic linear imaging phase. Both phases require that acoustic measurements taken by the IM nodes be time-synchronized. Several time synchronization algorithms for wireless sensor network have been developed in the literature [25]. For the purposes of this work, we do not make a hard choice on a particular algorithm. However, for practical reasons, it is recommended that a low-energy synchronization algorithm be used. For example, the clock synchronization algorithm developed by our research group in [26] is a suitable choice.

#### 1) Phase I: Active Acoustic Probing

The purpose of this phase is to discover the topology of the monitored GDN. In this phase, the IM nodes implement an extended version of the APR method in a round robin manner according to the following protocol:

1) Each IM node $i$ (Node $i$) in a cluster of $K$ nodes (i.e., $i = 1, \ldots, K$) receives a time schedule from the cluster head node specifying two time frames: a measurement frame $T_M$ and a data collection frame $T_C$. The frames are divided into time slots each assigned to a single sensing node. $T_M = \{t_m^1, t_m^2, \ldots, t_m^K\}$, $T_C = \{t_c^1, t_c^2, \ldots, t_c^K\}$.
2) In $t_m^i$, the following steps are performed:
   a. Node $i$ (i.e., active node in $t_m^i$) broadcasts a START beacon to every Node $j$ in the cluster, $j \neq i$. The START beacon includes a timer value that indicates recording start time of current time slot.
   b. At the expiration of the timer, Node $i$ injects an acoustic pulse into its pipeline terminal and records the pulse and its reflections. This signal is denoted $r_i[n]$. Simultaneously, every Node $j$, $j \neq i$ in the cluster starts recording to capture the acoustic pulse as transmitted to their terminals through the pipeline. The signal recorded by Node $j$ is denoted $t_{ij}[n]$.
   c. Step b can be repeated multiple times and the averages $\bar{r}_i[n]$ and $\bar{t}_{ij}[n]$ of repeated measurements are taken to cancel noise.
   d. Node $i$ broadcasts a STOP beacon indicating the end of measurement time for current time slot.
3) Step 2 is repeated for each $t_m^i$ in $T_M$ giving every node in the cluster a turn to be an active node.
4) At this point, each Node $i$ has one pulse reflection signal $\bar{r}_i[n]$ and $K-1$ pulse transmission signals $\bar{t}_{ji}[n]$ corresponding to the other $K-1$ nodes.
5) The data collection frame $T_C$ starts: During each time slot $t_c^i$, Node $i$ sends its $\bar{r}_i[n]$ and $\bar{t}_{ji}[n]$s to the cluster during.

At the end of step 5, the cluster head collects $K$ pulse reflection signals $\bar{r}_i[n]$ and $(K^2 - K)$ pulse transmission signals $\bar{t}_{ji}[n]$ from all nodes, and processes the signals to infer the distances between IM nodes and the positions of pipe joints (e.g., L-, T-, or cross-junctions) between them and subsequently discovering the topology of the pipeline network. The key information are in the $\bar{r}_i[n]$ signals which include pulse reflections corresponding to other IM node terminals (terminal reflections) as well as reflections corresponding to pipe joints (joint reflections) that appear between the injected pulse and the first terminal reflection. These pulse reflections are identified using information extracted from the $\bar{t}_{ji}[n]$ signals. This is illustrated in the experiment shown in Figure 3, where three IM nodes (IM1-IM3) performed the acoustic probing protocol described above on a simple pipeline network filled with air (cluster head not shown). The $\bar{r}_1[n]$, $\bar{t}_{12}[n]$, and $\bar{t}_{13}[n]$ measurements recoded during $t_m^1$ are plotted against acoustical distance ($c \times t$), where $c = 343 \; m/s$ is the speed of sound, in Figure 4. The signals were de-noised using 5-level DWT with a Symlet 6 wavelet function.

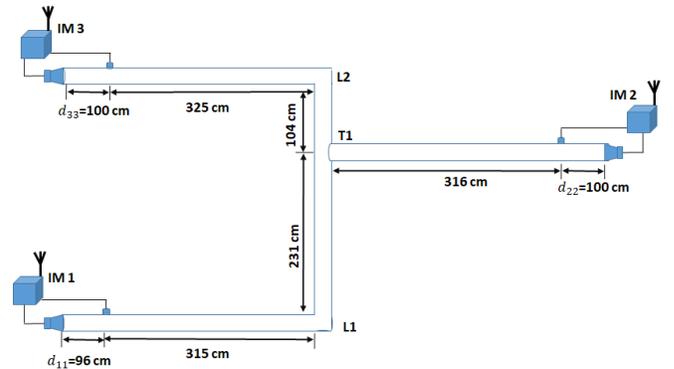

**Figure 3 - Active acoustic probing experiment in pipeline network with three sensing nodes**

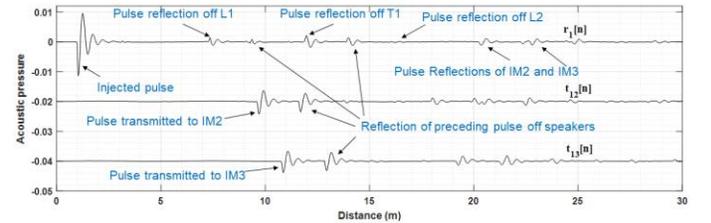

**Figure 4 - Reflected and transmitted pulse signals measured during $t_m^1$ (when IM1 is active)**

In Figure 4, the terminal and joint reflections are labeled in blue text on $\bar{r}_1[n]$. The other pulses are "speaker reflections", which are secondary pulse reflections that appear in the signal because reflected pulses propagating towards an IM terminal are reflected off its speaker and picked again by its mic. The distance between a pulse and its speaker reflection is twice the distance the mic-speaker distance of an IM node, which is denoted $d_{ii}$ for IM$i$ in Figure 3. The mic-speaker distance $d_{ii}$ is necessary to prevent interference between pulses and their speaker reflection and should be chosen to satisfy the condition: $2d_{ii} > c\Delta t$, where $\Delta t$ is the width of the injected pulse and $c$ is the speed of sound. Since this distance is a known design parameter, speaker reflections in $\bar{r}_i[n]$ and $\bar{t}_{ji}[n]$ can be readily identified. After those are identified in $\bar{r}_1[n]$, the terminal reflection in $\bar{r}_1[n]$ are determined by estimating the acoustical distance (i.e., distance through pipes) between IM1 and the other IM nodes. These distances, $\{d_{1j} : j = 2,3\}$, are estimated from the cross-correlation function between $\bar{r}_1[n]$ and $\bar{t}_{1j}[n]$ signals as follows:

$$d_{1j} = \frac{c}{F_s} \times \arg\max_m \left|\hat{R}_{\bar{r}_1 \bar{t}_{1j}}[m]\right| + d_{jj} \quad (2)$$

where

$$\hat{R}_{\bar{r}_1 \bar{t}_{1j}}[m] = \sum_{n=0}^{N-m-1} \bar{t}_{1j}[n+m] \bar{r}_1[n] \quad (3)$$

for $m, n = 0, .. N-1$. and $d_{jj}$ is the mic-speaker distance of node IM$j$.

Once the terminal reflections in $\bar{r}_1[n]$ are determined, any pulses that are sandwiched between the injected pulse and the first terminal pulse are considered joint reflections. Joint reflections can also be identified by their reversed polarity because they introduce a positive change in the pipe's cross-sectional area (see equation (1)). In Figure 4, it is noticed that terminal reflection pulses have the same polarity of the injected pulse, while joint reflection pulses have an opposite polarity. The distances between IM1 and those joints can be estimated by finding the time shifts of negative peaks in the auto-correlation function of $\bar{r}_1[n]$, which can be calculated using the same formula in (3) and replacing $\bar{t}_{1j}[n]$ with $\bar{r}_1[n]$.

By repeating the same steps above to $\bar{r}_2[n]$ and $\bar{r}_3[n]$, the cluster head node can confirm the position of all joints in the network and identify their types. Figure 5 shows these distances labeled on all $\bar{r}_i[n]$ and $\bar{t}_{ji}[n]$ signals collected in the experiment of Figure 3. The additional information extracted from the $\bar{t}_{ji}[n]$ gives this method an advantage over the traditional APR methods described in the previous section in terms of identifying pipeline features.

Assuming all pipe joints are right-angled, it is possible to identifying the type of pipe joints by reasoning from the distances extracted in Figure 5. However, such solution becomes intractable for complex pipeline network. A better solution for joint type identification is to cast the problem as a classification problem, in which the magnitude of the joint reflection pulse, which is proportional to the change in the cross-sectional area at the joint (see equation (1)), is used in identifying the type of pipe junction. In this work, we considered five types of pipe junctions that are similar to joints commonly used in gas distribution pipelines: 1"-to-1" L-bend, 1"-to-1" T-branch, 1"-to-1.5" T-branch, 1"-to-2" T-branch, and 1"-to-1" cross junctions. We used the ratio of reflected to injected pulse magnitude as input feature. This ratio defined as follows:

$$R_E = \frac{\frac{1}{t_4 - t_3} \sum_{n=t_3}^{t_4} r_i^2[n]}{\frac{1}{t_2 - t_1} \sum_{n=t_1}^{t_2} r_i^2[n]} \quad (4)$$

Where the pair of time indices $\{t_1, t_2\}$ and $\{t_3, t_4\}$ delineate the beginning and end of the injected and reflected pulses, respectively. A training data set was collected in a series of experiments where the distance between the IM node and the joint was varied. The data for each class was then fitted to an exponential curve of the form $R_E = \alpha e^{\beta d}$, where d is the distance, and new input data were classified based on their Euclidean distance to those curves. The training data and fitted are shown in Figure 6. Test data form 9 different experimental measurements in pipes up to 7 meter long showed that the classifier can accurately predict the type of the pipe joint even when acoustic transducers of different models are used.

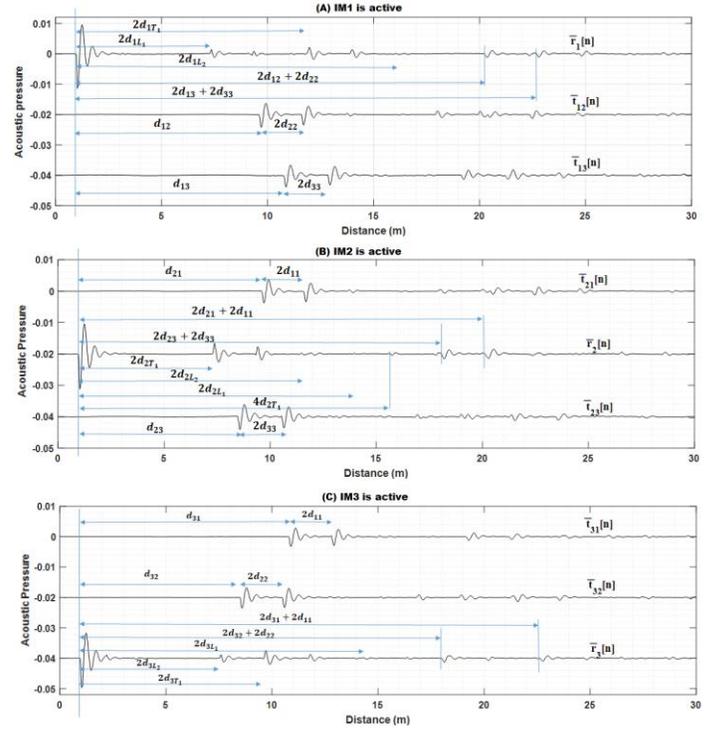

**Figure 5 – Annotated pulse reflection and transmission signals recorded in the experiment of Figure 3 when: (A) IM1 is active, (B) IM2 is active, and (C) IM3 is active.**

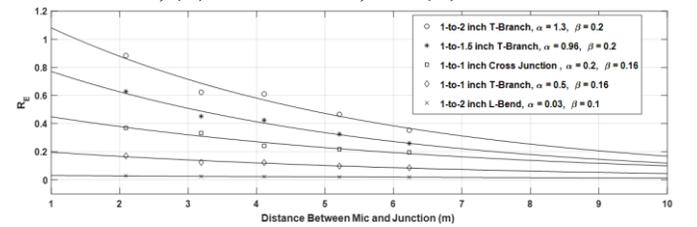

**Figure 6 - Exponential ($\alpha e^{\beta}$) curve fitting of $R_E$ measurements as function of distance**

*2) Phase II: Passive Acoustic Linear Imaging*

Once the active acoustic probing phase is completed and the pipeline network topology is discovered, the IM nodes start the passive imaging phase. In this phase, the sensing nodes listen synchronously to acoustic activity in the pipeline for a period of time and perform a modified version of the Time-Exposure Acoustic (TEA) imaging algorithm described in [27]. Similar in principle to time-exposure photography, the TEA algorithm is intended to image low-SNR continuous acoustic sources and is based on the assumption that signals of such sources are spatially correlated unlike background noise. Therefore, by coherently summing the correlated signal over time, the contrast of the source signal increases against the uncorrelated background noise. In the context of leak detection in GDNs, leak acoustic signals, particularly of small and distant leaks, have low SNR and are continuous in nature. We describe how the IM nodes apply the TEA algorithm in the following.

Let each IM node $N_i$ have a location in the Cartesian space $l_i = (x_i, y_i, z_i)$ and the acoustic signal recorded by $N_i$ be $p_i(l_i, t)$. To image the pipeline, the pipeline is first pixelated along its length into image points each with a location in space $l = (x, y, z)$. Each image point is assigned an image intensity value using the following image function:

$$I(l) = \sum_{i=1}^{N} |\hat{p}_i(l_i, |l - l_i|/c) e^{K|l-l_i|/c}|^2 \quad (5)$$

Where $\hat{p}_i(l_i, t) = p_i(l_i, t) * F(t)$ and $\hat{p}_i(l_i, |l - l_i|/c)$ is the signal backpropagated in space to the point $l = (x, y, z)$, $F(t)$ is a filter applied to the signal, the $e^{K|l-l_i|/c}$ is added to compensate for wave attention in the pipe, $c$ is the speed of sound, and $K$ is the wave attenuation coefficient. The signals in (5) are squared because acoustic measurement signals are zero-mean. However, this could create a large DC bias [27]. To avoid that, the following unbiased estimator is used:

$$I^{(j)}(l) = \left|\sum_{i=1}^{N} \hat{p}_i^{(j)}(l)\right|^2 - \sum_{i=1}^{N} \left|\hat{p}_i^{(j)}(l)\right|^2 \quad (6)$$

Where $\hat{p}_i^{(j)}(l) = \hat{p}_i(l_i, |l - l_i|/c) e^{K|l-l_i|/c}$ is the backpropagated signal used at the $j^{th}$ realization of the image. The final image $\bar{I}(l)$ can then be obtained by averaging over all realizations:

$$\bar{I}(l) = \frac{1}{M}\sum_{j=1}^{M} I^{(j)}(l) \quad (7)$$

To apply the TEA algorithm to real data, the experiment shown in Figure 7 was conducted. A pre-recorded leak signal was replayed using a speaker at 260 cm from IM node1. The leak signal was recorded by generating a hole next to mic in a pipe containing pressured air at 4 psi.

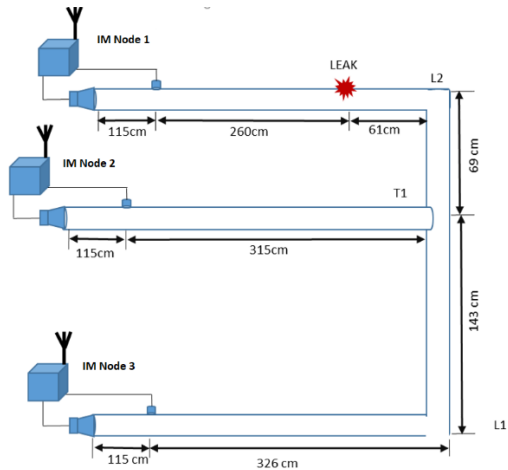

**Figure 7 – Experiment using three IM nodes to apply the TEA algorithm in leak detection and location.**

The linear image resulting from applying the TEA algorithm to the signals acquired by the three IM nodes is shown in Figure 8. The image was produced using equation (7) with 100 realizations, where each realization was produced by applying equation (6) to signal segments of length = 0.4 seconds. The linear image shows a clear peak at the location of the leak. It also shows higher image values throughout the branch of IM Node 1 (Branch 1) compared to the other branches. This is because the leak signals recorded by IM Nodes 2 and 3 are fully correlated throughout Branch 1. In other words, to IM Nodes 2 and 3, the leak signal appears to be coming from the joint T1 and the signal recorded by IM Node 1 is necessary to pin down the exact location of the leak.

It is noticed that the linear image in Figure 8 contains multiple secondary peaks despite averaging the signals over time to cancel incoherent background noise. These secondary peaks are caused by in-pipe reflections in the $p_i(l_i, t)$ signals. They can be suppressed using an echo-canceling filter for $F(t)$ in equation (5). One way to do that is using an ideal echo-canceling filter of the form [28]:

$$H(z) = \frac{1}{1 - \alpha z^{-\tau}}$$

This filter assumes that $p_i(l_i, t)$ has the form $p(t) = s(t) + \alpha s(t - \tau)$ where $s(t)$ is the leak signal and $\alpha s(t - \tau)$ is its speaker reflected version. The parameters $\alpha$ and $\tau$ can be estimated from $\bar{t}_{ji}[n]$ during the active probing phase, as $\tau$ is directly related to mic-speaker distance of the IM node (See $d_{ii}$ in Figure 5). However, this filter only cancels the speaker reflections. A more efficient way to suppress reflections is using an adaptive filter as shown in Figure 9. Using this method, IM nodes can utilize their wireless connections to train each other adaptive filters. In Figure 9, IM Node $j$ injects a sequence of orthogonal acoustic signals (e.g., sinusoids) $x[n]$ into the pipe, while IM Node $i$ records it as $x[n] * h[n]$. Simultaneously, IM Node J sends $x[n]$ wirelessly to IM Node I, which uses $x[n]$ to train the filter echo canceling filter $\hat{h}[n]$. The filter can be further trained using training signals from other IM nodes in the cluster.

The pipeline linear image resulting from applying the TEA algorithm to signals filtered with echo-canceling filter is shown in Figure 10 along with the linear image from Figure 8. Suppressed secondary peaks are marked with arrows.

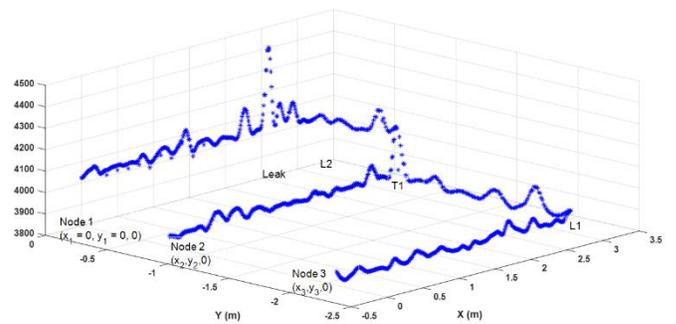

**Figure 8 – Linear image of the pipeline produced using the TEA imaging algorithm ($c = 343 \, m/s$, $K = 0$, and $F(t) = \delta(t)$)**

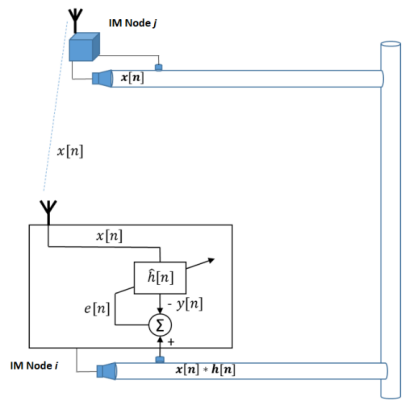

**Figure 9 – Cross-IM-node adaptive filter training for echo canceling.**

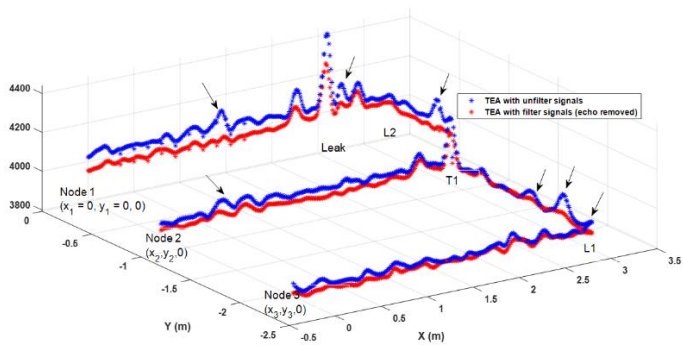

**Figure 10 - Comparison of TEA linear image using raw acoustic signals (blue) vs. signals filter with echo canceling filters (red).**

Finally, we note that this method can incorporate wireless shut off valves that stop gas flow in leak affected segments of a GDN. The method can also be extended to monitor water distribution networks.